\documentclass[11pt,a4paper,final]{article}
\usepackage[pdftex]{graphicx}
\pdfoutput=1
\usepackage[centertags]{amsmath}
\usepackage{amssymb}
\usepackage{mathrsfs}
\usepackage{gensymb}
\usepackage{color}
\usepackage[square, numbers]{natbib}
\usepackage[a4paper]{geometry}
\usepackage{psfrag} 
\usepackage{setspace}

\usepackage[pdftitle={}, 
  pdfauthor={Michael T. Morris-Thomas and Sverre Steen},
  pdfkeywords={fluid-elastic instability, flutter, static divergence,
    optical tracking}, colorlinks=true,citecolor=red]{hyperref}

\newcommand{\Ren}{\text{Re}}
\newcommand{\order}{\mathcal O}
\newcommand{\A}{\mathscr A}

\begin{document}

\title{\textbf{The response of a tensioned flexible sheet immersed in
    parallel flow}}
\author{Michael T. Morris-Thomas$^{\dag}$ and Sverre Steen\\
  \small{Department of Marine Technology,
    Norwegian University of Science and Technology,}\\
  \small{Otto Nielsens vei 10, Trondheim, NO-7491, Norway.
    $^{\dag}$E-mail: \texttt{michaelmorristhomas@gmail.com}}
}
\date{}%

\maketitle

\begin{abstract}
  This paper explores the fluid-elastic response of a cantilevered
  flexible sheet in the presence of uniform airflow. The leading edge
  of the sheet is clamped, while at the trailing edge, in-plane
  tension is applied to provide additional rigidity to the sheet's
  small but finite bending stiffness. We outline a series of
  experiments performed in a wind tunnel with the purpose of examining
  fluid-elastic instabilities. In particular, we examine the role of
  in-plane tension induced rigidity and how it influences static
  divergence and convected wave instabilities. The flow is
  characterised by Reynolds numbers of order $10^5$-$10^6$ and we
  specifically examine a sheet with an aspect ratio of $L/l=1.33$.  A
  unique aspect of this present work, is the direct measurement of the
  sheet's three-dimensional displacement through an optical tracking
  method with a grid of passive markers placed on the sheet
  surface. We show the evolution of the sheet surface from stability,
  through to divergence, and then finally into flutter. The frequency
  composition of the flutter event shows higher harmonic components
  that suggest significant nonlinearities. Tension induced rigidity
  plays a crucial role in the response of the sheet to the fluid in
  both postponing and suppressing instabilities.
\end{abstract}

\onehalfspacing

\section{Introduction}

The response of a flexible surface immersed in a moving fluid is a
ubiquitous fluid-structure interaction problem. It can be observed in
everyday life from the motion of a leaf fluttering in a mild breeze,
to the locomotion of insects, birds and waterborne animals, to more
abstract applications like snoring and marine vehicles. Despite this
ubiquity, much is still to be understood because the inherent
fluid-elastic coupling of the problem poses certain difficulties in
determining fluid-elastic stability characteristics. Consequently,
this limits our understanding of a flexible surface's response under large
displacements, the scaling of its fluid dynamic drag, three-dimensional
effects, and the role of both shear and oscillatory boundary
layers. In this present work we experimentally examine the behaviour
of a flexible sheet both before and after stability is lost with
particular emphasis on static divergence, convected wave instability
modes, and the influence of tension applied at the trailing edge of
the sheet on these instabilities.


A flexible sheet immersed in a flowing fluid can experience
instability when the incident fluid velocity breaches some critical
value \cite{Rayleigh1879}. The nature of this instability depends upon
competition between the sheet´s inertia, fluid momentum, and the
restorative influences of its flexural rigidity and axial or lateral
tension --- which can be supplied by either the boundary layer or by
external means. At the onset of instability, which can be caused
through any significant perturbation of the system, the manner in
which these characteristics of the system interact can result in
either a convected wave type of instability
\cite{Peskin2002,Dowell2003,Connell2007} or static divergence
\cite[]{PaidoussisV2}. We shall denote the velocity at which these
occur $U_s$ and $U_c$ for static divergence and convected waves
respectively. A convected wave type of instability is characterised by
dispersive waves that originate at the leading edge of the sheet and
propagate towards the trailing edge. This type of instability is often
referred to as flutter where $U_c$ is thought of as the critical
velocity for a global instability. On the other hand, in the event of
divergence, the sheet typically adopts an alternative mean position in
response to the flow that can be one of infinitely many possible mode
shapes. Typically, if divergence is to be realised then $U_s <
U_c$. In both instability regimes, this self-excited response can
adversely affect the sheets hydrodynamic performance, fluid dynamic
drag, and can lead to more chaotic behaviour --- see Connell and
Yue's \cite{Connell2007} Figure 23 for instance.

To simplify a description and aid in our understanding of the
fluid-elastic coupling of the problem, we can divide the nature of the
surrounding fluid flow into two regimes based on the sheet aspect
ratio $\mathscr A = L/l$ where $L$ and $l$ denote the sheet length and
width respectively. For $\mathscr A \gg 1$, we essentially have a long
slender sheet and the flow about it can be considered approximately
two-dimensional in the cross-flow direction. Here, flow separation at
the trailing edge is of little consequence compared to boundary layer
induced tension and fluid added mass effects. Such a characterisation
would encompass the performance of long slender bodies of more general
cross-section \cite{Lighthill1960} and in particular ribbons and
streamers \cite{Datta1975,Auman2001,Carruthers2005,LeHeLa2005}.

On the other hand, when $\mathscr A=\order(1)$, the characteristics of
the flow are far more complicated due to the three-dimensionality of
the sheet and the importance of flow separation and shear layers in
the form of an unsteady wake and downwash over the cross-flow
direction. In a strictly two-dimensional setting, the effect of an
unsteady wake on a sheet of infinte width has been considered both
theoretically \cite[]{Kornecki1976,Peskin2002,ArgMah05} and
experimentally in the illuminating soap film experiments of Zang and
associates in a post-critical regime \cite{Zhang2000}. However, in the
regime $\mathscr A=\order(1)$, the coupling of unsteadiness with the
effects of three-dimensionality has only recently been approached
numerically \cite[][]{Dowell2003} with some success at determining the
critical flow velocity for a flexible sheet and its qualitative
behaviour after stability is lost.

For sheets comprising small flexural rigidity, the majority of
previous experimental studies concern the regime $\mathscr A \gg
1$. Here, the sheet is often mounted vertically
\cite{Taneda1968,Datta1975,LeHeLa2005} allowing gravity to provide
additional restorative restraint to flexure. On the other hand, it is
possible to mount a flexible sheet horizontally in flow with axial
tension \cite{Coene92}, lateral tension \cite{Watanabe2002a} or a
combination of both \cite{MMTSS09}. This approach is advantageous if
one wishes to quantify the effect of tension induced rigidity on
instability. In a recent experimental campaign, Morris-Thomas and
Steen \cite{MMTSS09} considered the regime $\mathscr A=\order(1)$ with
trailing edge tension providing additional rigidity both axially and
laterally to a flexible sheet. The authors demonstrated that axial
tension can temporarily postpone fluid-elastic instability, leading to
a reduction in fluid dynamic drag, and consequently suppressing the
expected rise in drag once stability is lost. However, although the
relationship between stability and drag were examined, no direct
measurement of the displacement of the sheet surface was performed. At
present we lack quantitative information concerning the
three-dimensional performance of a flexible sheet, how in-plane
tension affects the growth and nature of fluid-elastic instabilities,
and the consequences of these to the fluid dynamic drag. We are not
aware of any experimental studies that attempt a direct measure of the
three-dimensional displacement of a flexible sheet in flow. We intend
to address this situation here and present some results from our
recent experimental study.


We investigate the fluid-elastic response of a cantilevered sheet of
low flexural rigidity $B$ and mass per unit area $m$ immersed in
uniform airflow defined by velocity $U$. The sheet is mounted
horizontally in the flow and tension is supplied at the trailing
edge. We describe a set of experiments that were performed on the
configuration in a wind tunnel. A test matrix comprising Reynolds
numbers of order $10^5$-$10^6$, sheet aspect ratios $\mathscr A =
1.33$-$2.0$, and a variation in trailing edge tension was
considered. The displacement of the sheet surface was monitored via an
optical tracking system whereby a linear grid of approximately 100
retro-reflective markers was placed on the sheet surface. The
utility of this approach allows us to study the evolution of the sheet
surface from a stable into an unstable state without intruding on the
fluid flow. In this paper, we examine the trajectories provided by
these passive markers for one particular sheet $\mathscr A =
1.33$. This provides a unique view into static divergence and
convected wave type instabilities acting over the surface of a
sheet. Furthermore, we examine these instabilities in the presence of
a parametric variation of tension induced rigidity acting on the
trailing edge of the sheet. The paper is organised as follows: a
description of the experimental campaign undertaken is given in \S
\ref{sec:exp}; a discussion of the experimental results in terms of
flutter and divergence is provided \S \ref{sec:dc}; and finally,
conclusions and suggestions for future work are outlined in \S
\ref{sec:conc}.

\section{Experimental Campaign}\label{sec:exp}

Experiments were conducted in the low speed wind tunnel of the
Aerodynamics Laboratory at the Norwegian University of Science and
Technology. The tunnel comprises an available test section of 2.7m
$\times$ 1.83m and allows fluid velocities up to 30ms$^{-1}$ with a
turbulence intensity of less than 5\%. The experimental set-up and
testing matrix is in some respects an extension of that described by
Morris-Thomas and Steen \cite{MMTSS09}. However, while this expanded
experimental campaign employs a smaller test matrix, it encompasses
direct measurements of the sheet displacement under flow.

\begin{figure}[t]
  \centering
  \includegraphics[scale=0.12]{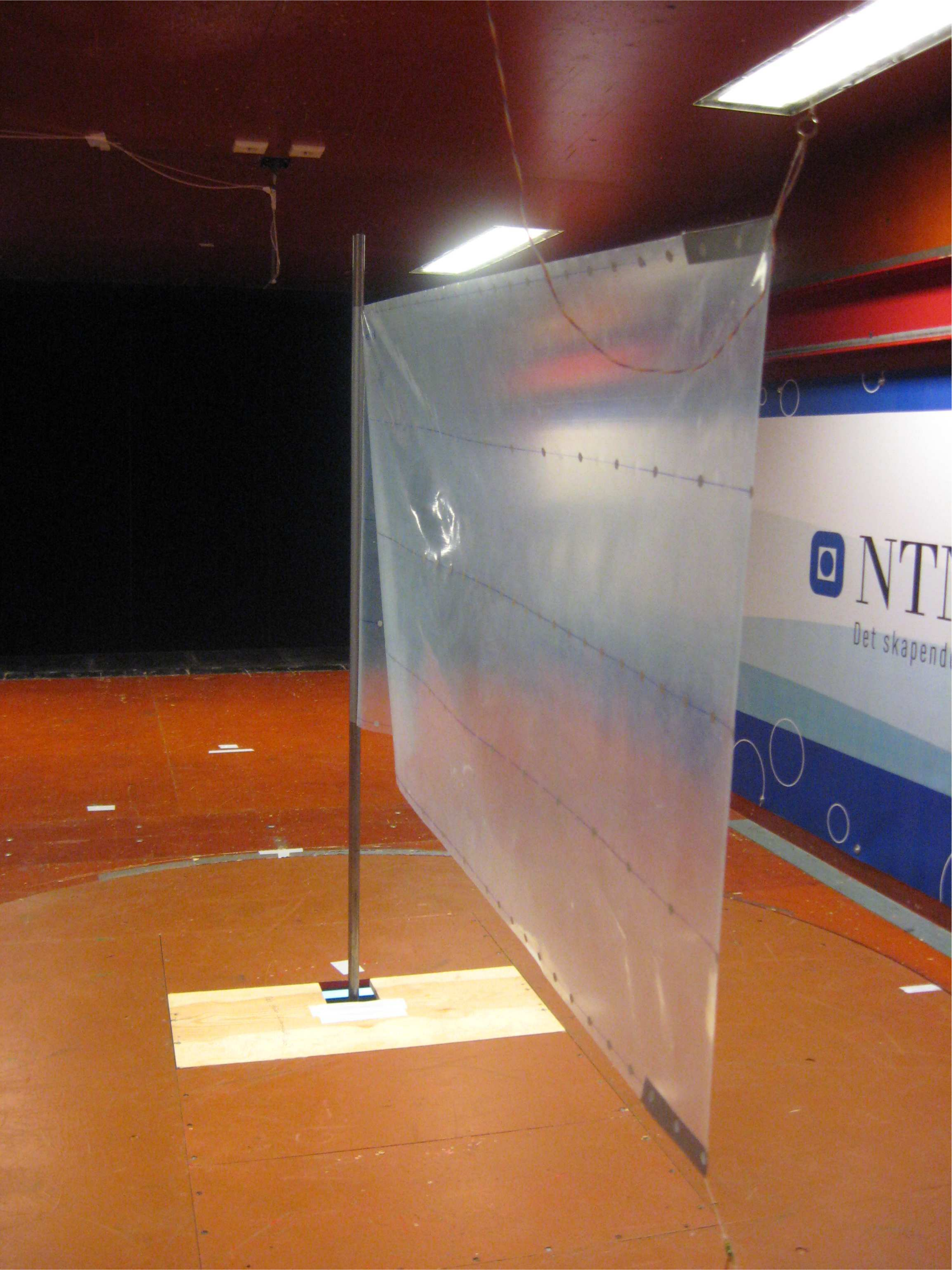}
  \caption{The experimental set-up in the wind tunnel. The
    retro-reflective markers are pictured in grey over five columns
    along the sheet surface. Although not pictured, the two ProReflex
    cameras are located to the left of the image outside the wind
    tunnel. Two nylon strings are observed at the trailing edge which
    were used to apply tension.}
  \label{fig:setup}
\end{figure}

A transparent Polyethylene sheet of thickness $h=0.15$mm and width
$l=75$cm was mounted horizontally along the centre-line of the test
section with a clamped leading edge and free trailing edge. The set-up
is illustrated in Figure \ref{fig:setup}. The bottom of the sheet was
located 54cm above the base of the wind tunnel avoiding the developing
turbulent boundary layer along the wind tunnel floor and walls. The
pertinent characteristic of the sheet include a small flexure
rigidity $B=59.7 \times 10^{-6}$Nm$^2$m$^{-1}$ to allow the
observation of flexural waves and a small mass per unit area
$m=0.14$kgm$^{-2}$. At the trailing edge of the sheet, tension was
applied via nylon strings attached to the top and bottom vertices of
the sheet (cf. Figure \ref{fig:setup}). Each string left the trailing
edge at a known angle $\theta$ and travelled through a pulley system
that directed it underneath the wind tunnel where it was attached to a
known weight providing a combined pre-tension $T$.

A typical test run involved gradually increasing the fluid velocity in
the wind tunnel over increments of approximately $0.2$-$0.5$ms$^{-1}$
until instability of the flexible sheet was observed. Testing then
continued up to fluid velocities of approximately
$U_c+2$ms$^{-1}$. The testing matrix adopted here is summarised in
Table \ref{tab:parameters} where we note a parametric variation of the
sheet's aspect ratio $\mathscr A = L/l$, the magnitude of combined
tension and the angle at which it was applied to the trailing edge. The
measurable quantities of interest to us are the fluid dynamic drag
experienced by the sheet, the mean and dynamic tension at the trailing
edge and the displacement of the sheet surface.

\subsection{Apparatus}

In order to acquire the displacement of the sheet surface we employed
optical tracking. The advantage of such a system is that it allows a
non-intrusive approach to measuring the displacement of the sheet
surface without disrupting the fluid flow. The flexible sheet was
fitted with a grid of retro-reflective markers each comprising a 4mm
diameter. These passive markers were spaced at 50mm intervals in the
flow direction and 180mm in the cross-flow direction. This provided a
grid of 100 markers with five columns for a sheet length of 1m. These
markers can be observed in Figure \ref{fig:setup}. The displacement or
trajectories of these markers in space was captured with two ProReflex
motion capture cameras.  These cameras were mounted in a vertical
alignment outside the wind tunnel approximately 2m from the
centre-line of the sheet surface. Vertical alignment, such that the
top and bottom cameras were positioned above and below the top and
bottom edges of the sheet respectively, was chosen to avoid refraction
errors introduced by the perspex panels of the wind tunnel and sheet
surface itself. The origin of the $(x,y,z)$ co-ordinate system was
located at the bottom of the leading edge of the sheet. The positive
$x-$axis is directed down the sheet length and the positive $z-$axis
defines the location of the leading edge. Dynamic calibration of the
optical tracking system was performed approximately every 4 hours of
testing.

In addition to the sheet displacement, we also recorded the external
tension applied to the trailing edge and the fluid dynamic drag
experienced by the sheet. For the tension, a ring type strain gauge
was located on each nylon string line immediately behind the trailing
edge. The fluid dynamic drag was measured by using a three component
force balance situated under the centre of the test section. The
leading edge of the sheet was attached via nylon tape to a 1.43m long
vertical pole comprising an elliptical cross-section ($a=4$cm,
$b=2$cm). This pole extended underneath the wind tunnel and was
attached to the force balance allowing a direct measure of the fluid
dynamic drag. Previous testing utilising this pole \cite{MMTSS09}
showed that its drag coefficient is $C_D=0.59$ and therefore, its
contribution to the overall drag, although minimal, can be subtracted
from the in-line force measurements at the force balance. In doing so,
and after further subtracting the axial tension, provides us with a
measure of the drag experience by the sheet.

The incident flow velocity was measured with a Pitot-static tube
located 1m upstream from the leading edge position of the sheet. The
fluid properties that we adopt for the analysis are based on a
measured air temperature of 20.3\degree C implying
$\rho=1.205$kgm$^{-3}$ and $\nu=0.150$cm$^2$s$^{-1}$. Measurements
from the Pitot-static tube, force balance and strain gauges were
acquired through a HBM instrumentation amplifier with a wireless
connection to a notebook computer running Catman data acquisition
software. The measurements from all sensors were digitised at 200Hz and
pre-processed through a fifth-order low-pass Butterworth filter with a
cut-off frequency of 40Hz. The dominant frequencies of the system, and
associated harmonics of interest to us, were all less than
approximately 25Hz.

\subsection{Post-processing}

The digitized images were processed with Qualisys\footnote{see
  \texttt{http://www.qualisys.com} for further details regarding the
  motion capture system and software.} motion tracking software to
extract the trajectory $M_n(x_i,y_i,z_i,t_i)$ of each marker $n$ in
the global co-ordinate system over time $t=i\Delta t$ where $i$ is an
integer. In selecting $\Delta t$, the channel capacity of the motion
capture cameras forced us to compromise frame rate to allow a suitable
amount of passive markers for the sheet. Consequently, choosing an
upper bound of $N=100$ markers provided us with a maximum frame rate
of $f=60Hz$. Furthermore, the available buffer of the cameras meant
that we could safely record for time windows of length $T_L = 58$s
before overflow resulted in a loss of markers and hence
trajectories. This time window is further shortened should rouge
markers be detected by the cameras. To avoid such complications we
chose a frame rate of $f=50Hz$ and operated over time windows of $T_L
= 50$s.

With windowing taken into account, each element from the test matrix
resulted in approximately 10-17 data sets of constant flow velocity
comprising trajectories for around 90-100 markers. The trajectory data
was then exported into a standard ASCII format for post-processing in
MATLAB. Post-processing revealed several rouge markers introduced
through artificial light sources caused by reflections through the
perspex wind tunnel walls and the sheet surface. The positions of
which were subsequently identified and these markers removed from each
data set.    

\section{Results and Discussion}\label{sec:dc}

\begin{figure}[t]
  \centering
  \includegraphics{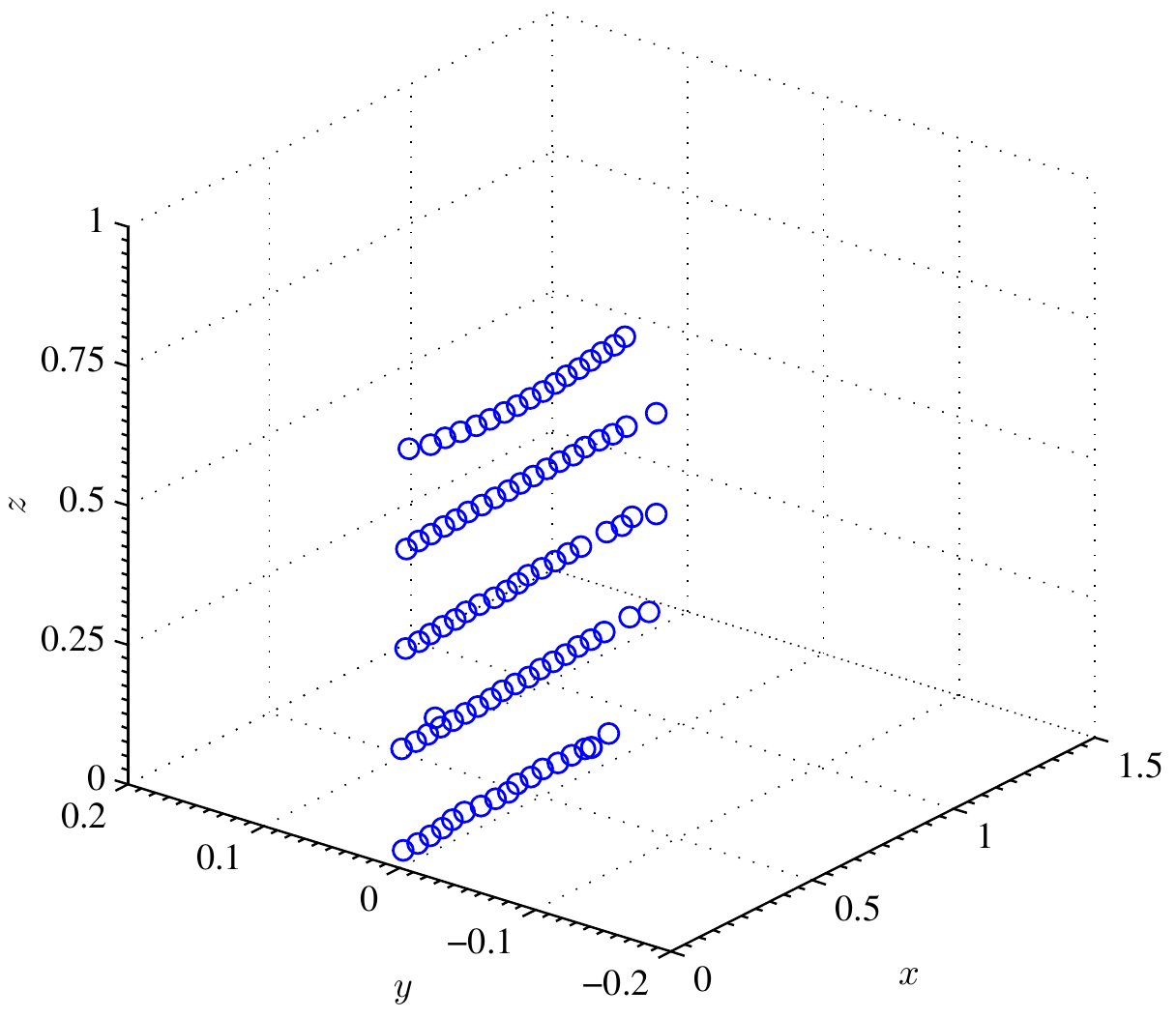}
  \caption{Markers positions on a perfectly stable sheet of length
    L=1m. These results correspond to an incident flow velocity of
    U=2.37ms$^{-1}$ with a combined tension of T=7.95N applied at the
    trailing edge of the sheet.}
  \label{fig:marks1}
\end{figure}

Typical passive marker positions on the surface of the flexible sheet
are shown in Figure \ref{fig:marks1} for $U=2.37$ms$^{-1}$ with a
combined tension of $T=7.95$N applied at 45\degree to the trailing edge. The
flow direction is from $x=-\infty$ --- this is implied for all results
presented hereafter. The surface of the sheet appears in the $(x,z)$
so that $x/L=0$ and $x/L=1$ correspond to the leading and trailing edges
respectively. This particular case illustrates a perfectly stable
sheet. When instability results we expect displacements of each marker
to occur across the sheet in the $(y,z)$ plane.

\begin{figure*}[t]
  \centering
  \includegraphics[width=\linewidth]{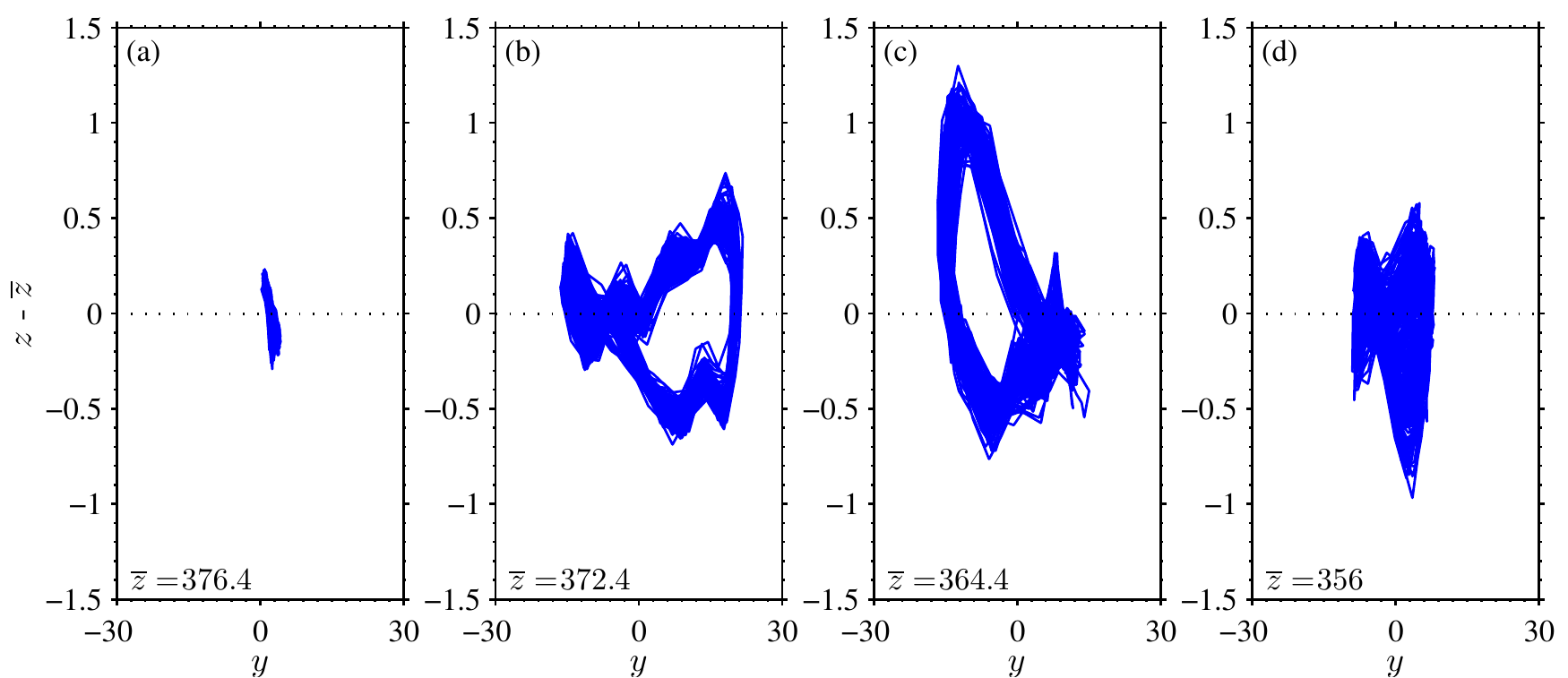}
  \caption{Trajectories of fours markers in the (y,z) plane along
    the centre-line of a sheet. These results correspond to the same
    configuration pictured in Figure \ref{fig:marks1} with the fluid
    velocity increased to U=7.26ms$^{-1}$. The sheet is now unstable to
    flutter. The marker positions are: (a) x$\approx$0.05; (b)
    x$\approx$0.35; (c) x$\approx$0.70; and (d) x$\approx$0.95. The
    insert on the bottom left hand corner of each axis gives the mean
    vertical component of the displacement. All units are in [mm].}
  \label{fig:marks2}
\end{figure*}

In contrast, we now show marker trajectories along the centre-line of
the same sheet after stability is lost to flutter. Figure \ref{fig:marks2}
illustrates the trajectories in time for four such markers in the
$(y,z)$ plane along the sheet's centre-line at the horizontal
positions $x\approx 0.05, 0.35, 0.70$ and $0.95$. We notice that the
displacement of the markers in the vertical direction is actually
quite small and $\order(1)$mm. The majority of the motion occurs along
the $y$-axis and, as such, the path of the trajectories in the centre
of the sheet are mildly elliptical with a large major axis. However,
at the extremities of the sheet, the clamped boundary condition at the
leading edge and the applied tension at the trailing edge, leads to a
more restrained horizontal motion. One problem with mounting a sheet
horizontally so that its leading edge is vertical is the possibility
of introducing asymmetry along the sheet surface. This is caused
through the weight of the sheet and we expect this to be more prevalent
under small magnitudes of applied tension. Figure \ref{fig:marks2}
actually shows the result of this where one can see the mean vertical
position of the markers decreasing as we progress further along the sheet
towards the trailing edge.

\subsection{Flutter instabilities}

\begin{figure}[t]
  \vspace{0.6cm}
  \centering
  \includegraphics{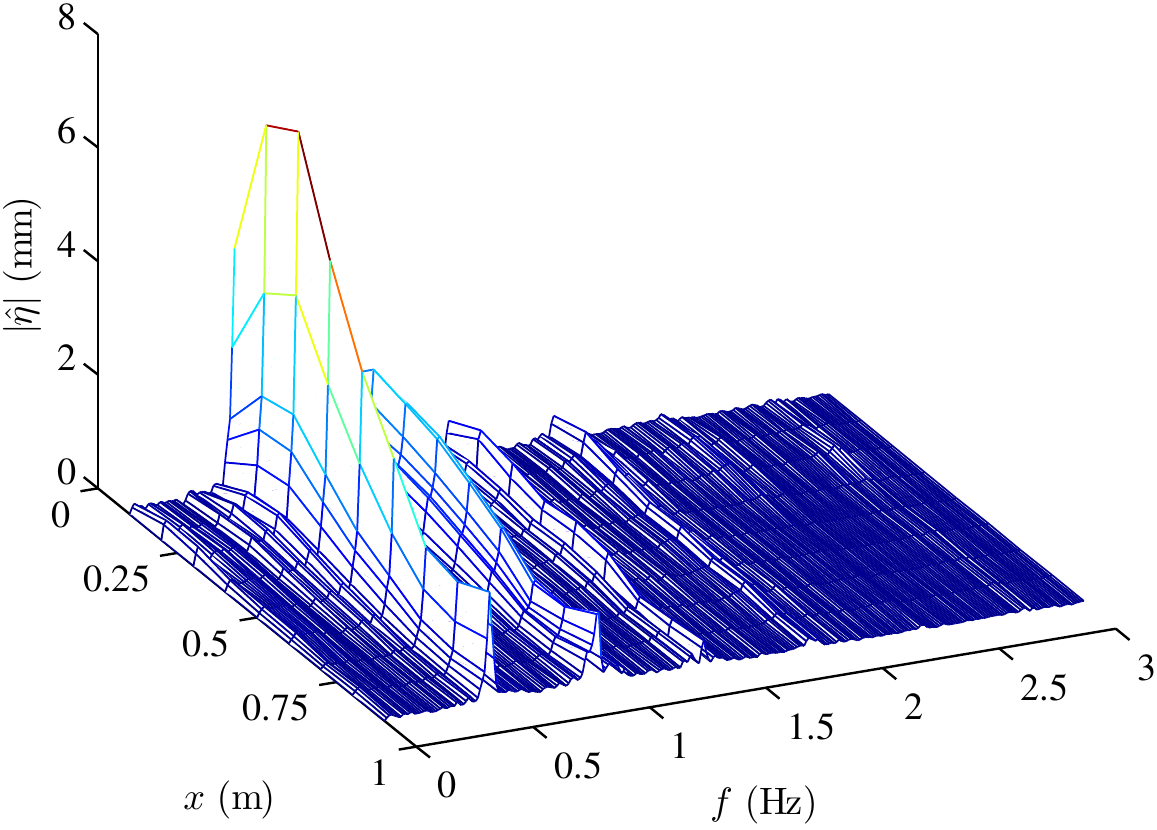}
  \caption{Spectral content of the displacement data for the markers
    along the centre-line of the sheet. The sheet is unstable at this
    incident flow velocity of U=2.42ms$^{-1}$. A combined tension of
    2.06N is applied to the trailing edge in this example.}
  \label{fig:ft}
\end{figure}

A good example of the frequency composition of a flexible sheet once
stability is lost to a convected wave instability is illustrated in
Figure \ref{fig:ft}. In this example we focus on a sheet of length
$L=1$m in an incident flow velocity of 2.42ms$^{-1}$ with a very moderate
combined tension of $2.06$N. Here, we have chosen to concentrate on the
centre-line of the sheet $z=0.375$ where values of $x=0$ and $1$ denote
the leading and trailing edges respectively. The $f$ axis corresponds
to the frequency content of the horizontal displacement of the sheet
measured in Hz. In frequency space, this is represented as $\hat \eta$
and obtained via a standard Fourier transform algorithm over the
signal length for each marker located along the centre-line.

We observe that once stability is lost to flutter, a dominant
frequency and associated phase locked harmonic components appear
(cf. Figure \ref{fig:ft}). The appearance of higher harmonic
components, the amplitudes of which are certainly not negligible in
relation to the first-harmonic, suggest nonlinearities in the
system. Presumably, these elastic nonlinearities result from nonlinear
sheet curvature which must clearly be important under large
displacements incurred through a flutter event. The measurements
suggest that the dominant frequency is $f_1=0.43$Hz with harmonics
visible up to $3 \times f_1$. Another interesting feature of the
results is that the global maximum of the convected wave amplitude
envelop occurs towards the leading edge of the sheet --- approximately
$x/L=0.25$. After which, the amplitude monotonically decreases towards
a local minimum in the vicinity of the the trailing edge. Although not
illustrate here, our results suggest that when the axial tension is
increased the position of this maximum amplitude is not significantly
affected.

\begin{figure}[t]
  \centering
  \includegraphics{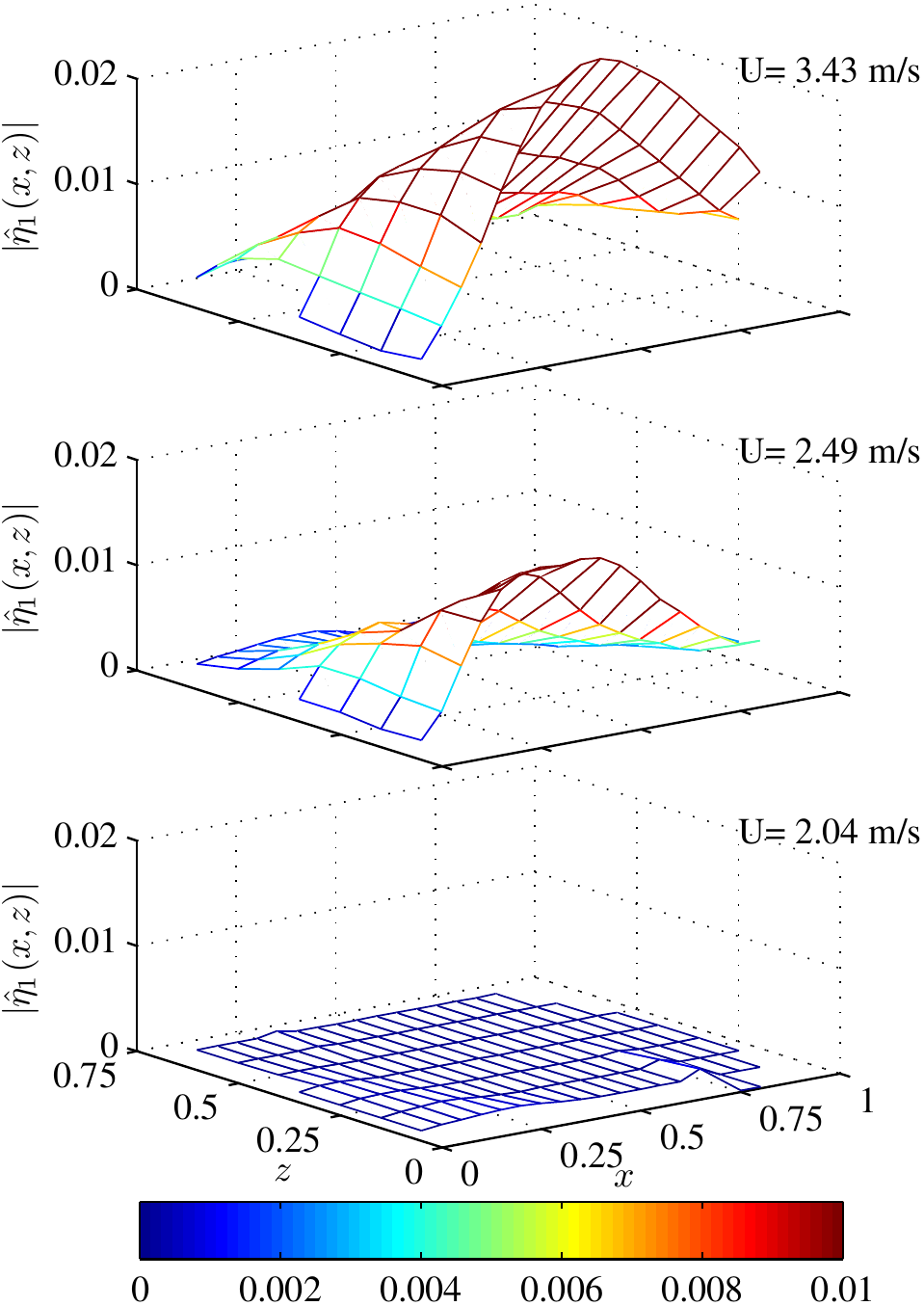}
  \caption{The magnitude of the first-harmonic component
    $|\eta(\omega_1)|$ of the sheet displacement. Its distributed is
    shown across the surface of the sheet located in the $(x,z)$
    plane. A combined tension of 2.06N is acting on the trailing
    edge. All units are in [m].}
  \label{fig:w1-4001}
\end{figure}

\begin{figure}[t]
  \centering
  \includegraphics{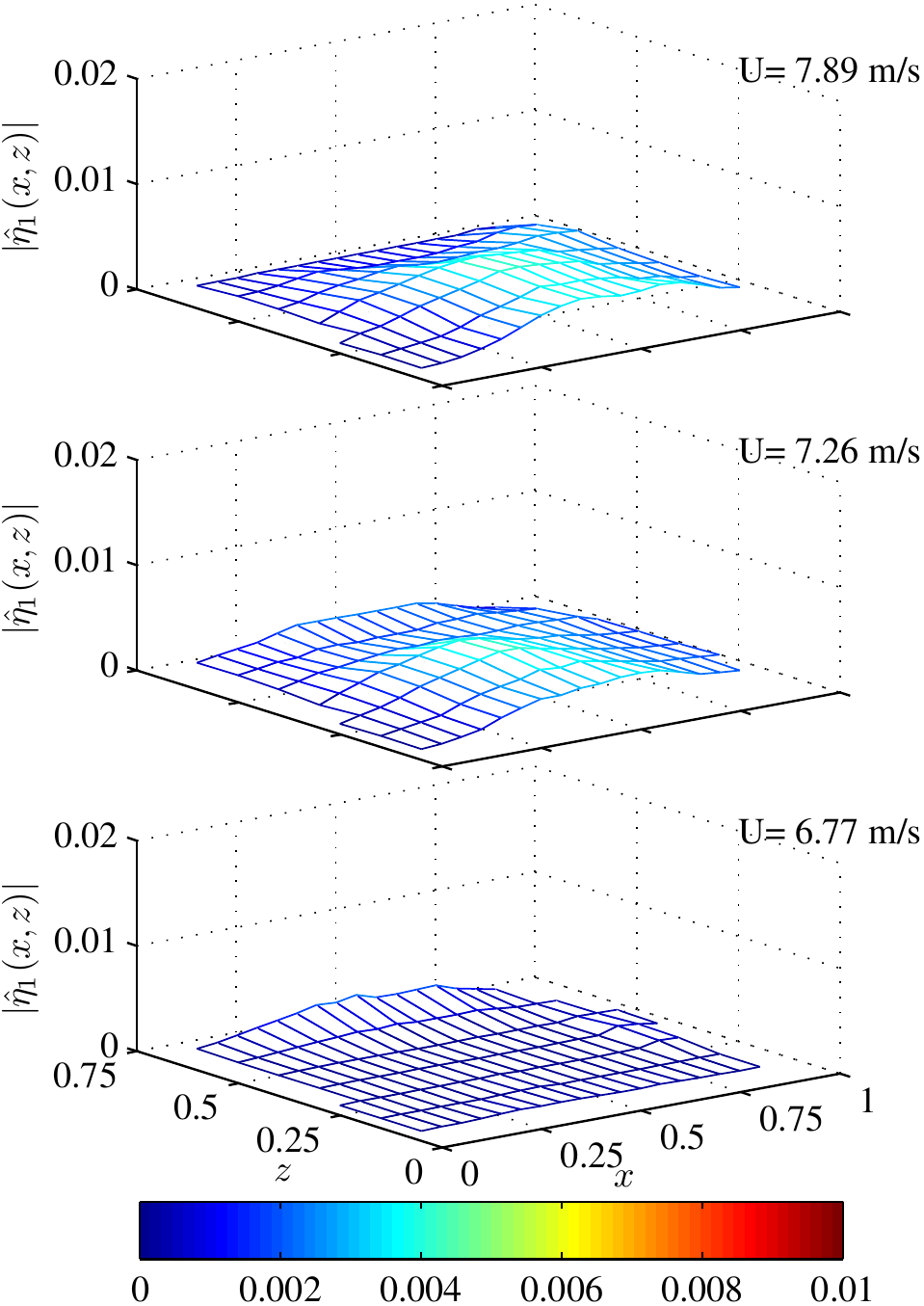}
  \caption{The magnitude of the first-harmonic component
    $|\eta(\omega_1)|$ of the sheet displacement. Its distributed is
    shown across the surface of the sheet located in the $(x,z)$
    plane. A combined tension of 5.98N is acting on the trailing
    edge. All units are in [m].}
  \label{fig:w1-4003}
\end{figure}

We now draw attention to the dominant or first-harmonic of the
instability and examine how this develops over a select range of fluid
velocities. To illustrate this, we present Figures \ref{fig:w1-4001}
and \ref{fig:w1-4003}. These plots show the spatial distribution of
the first-harmonic magnitude of the convected wave instability over
the $(x,z)$ plane. It is denoted $|\hat \eta_1(x,z)|$. This dominant
harmonic was obtained through a direct Fourier transform of the
$y(t)$-component of each marker on the sheet. A band-pass Butterworth
filter was then employed to extract the first-harmonic. The positions
of the markers in the $(x,z)$ plane were relatively invariant with
time compared to the $y$-component, so the mean positions obtained
from the time series' are adopted for the analysis here.

Once stability is lost, the first-harmonic appears quite noticeably
over the surface of the sheet. We can infer from Figure
\ref{fig:w1-4001} that stability is lost to flutter in the vicinity of
$2.04<U_c<2.49$ms$^{-1}$. While under under greater axial tension,
this range is translated to $6.77<U_c<7.26$ms$^{-1}$ (cf. Figure
\ref{fig:w1-4003}). At the top edge of the sheet, $z=0.75$, we notice
a small flutter event before the dominant convected wave instability
covers the entire sheet --- see Figure \ref{fig:w1-4003} for
$U=6.77$ms$^{-1}$. This feature is a localised standing wave and it was more
noticeable under higher values of tension at the trailing edge.

Due to the fact that the leading edge of the sheet was mounted
vertically, we can expect some asymmetry in the instability under
small in-plane tension. This is due to the weight of the sheet and
would explain the large amplitudes along the bottom edge $z=0$ once
stability is lost (see Figure \ref{fig:w1-4001}). The elastic stress
distribution, resulting from the applied tension, would presumably be
of much lower density in this region which renders it more susceptible
to stability loss.

With greater tension applied to the sheet we observe a smaller
amplitude of the first-harmonic and much more asymmetry over the sheet
surface (cf. Figure \ref{fig:w1-4003}). Interestingly, under high
tension the first-harmonic component is slightly visible before
$U=6.77$ms$^{-1}$ as shown in Figure \ref{fig:w1-4003}, however, its
amplitude is extremely small and surrounded by broadband noise.

One aspect of the experiments worth mentioning is the observation of
high frequency flutter in the immediate vicinity of the trailing
edge. Unfortunately, this was not acquired at sufficient quality by
the optical tracking system to yield any quantitative
results. However, we observe that this flutter is very localised and
does not results in any convective type of instability but rather
resembles a localised standing wave. The amplitude of this high
frequency event appears largest along the centre-line and does not
penetrate any further than approximately $0.05L$ into sheet from the
trailing edge.

\subsection{Static Divergence}

\begin{figure}[t]
  \centering
  \includegraphics{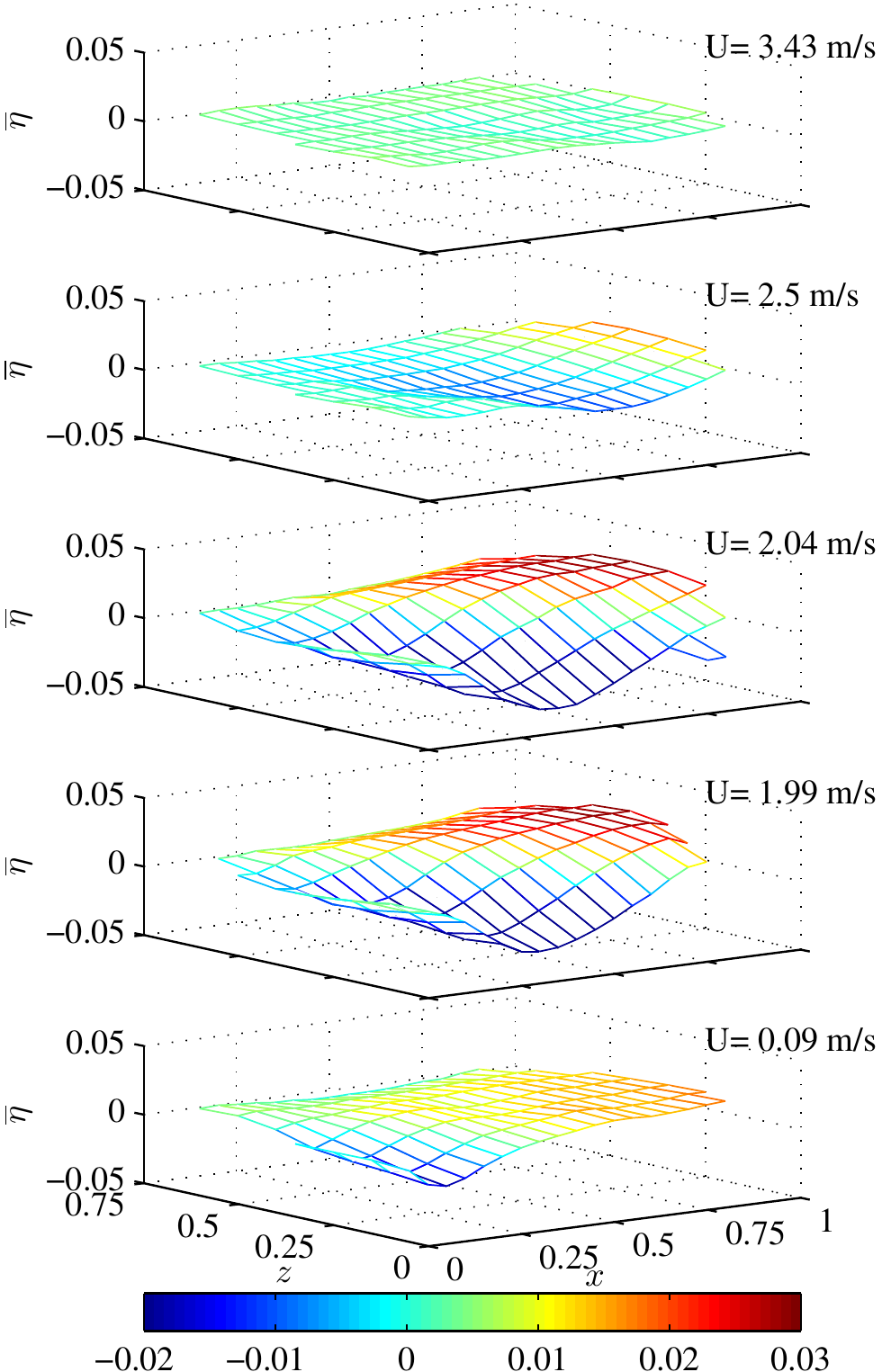}
  \caption{Magnitude of the zero-harmonic component across the surface
    of a sheet of length L=1m and width l=0.75m under a combined
    tension of T=2.06N.}
  \label{fig:w0-4001}
\end{figure}

We now examine the mean or zero-harmonic response of the sheet. This
should correspond to a static divergent type of instability. For a
sheet experiencing a combined tension of $2.06$N, the distribution of
the mean component over the sheet surface is illustrated in Figure
\ref{fig:w0-4001}. The mean component is denoted $\overline \eta$ and
the response is shown for a number of fluid velocities ranging from
$U=0.09$ms$^{-1}$ up to $U=3.43$ms$^{-1}$ which is beyond the point in
which flutter of the sheet was first observed (cf. Figure
\ref{fig:w1-4001}).

The appearance of a static instability is quite visible and the
results indicate that it occurs somewhere in the region of
$0.1<U_s<2$ms$^{-1}$. This range is somewhat lower than the flow
velocity at which a flutter instability mode first appears (cf. Figure
\ref{fig:w1-4001}). At $U=0.1$ms$^{-1}$ we observe a small mean
component in the vicinity of the leading edge along $x=0$. We suspect
this is unrelated to divergence and is merely caused by the weight of
the sheet under minimal trailing edge tension. However, it does offer
an irregularity or perturbation of the sheet profile that may promote
the onset of an instability.

Interestingly, once the sheet loses stability to flutter (cf. Figure
\ref{fig:w1-4001}) we notice a reduction in the mean component across
the surface of the sheet. In other words, divergence is being replaced
by a convected wave instability and it is quite possible that energy
is then transferred to the first-harmonic. In the overlapping region
between the two instabilities superposition would appear to be taking
place. However, after the convected wave instability is realised, the
static divergent mode is suppressed at $U=2.5$ms$^{-1}$ and it is then
almost inconsequential at $U=3.43$ms$^{-1}$ where it has been replaced
by flutter (see Figure \ref{fig:w0-4001}).

\begin{figure}[t]
  \centering
  \includegraphics{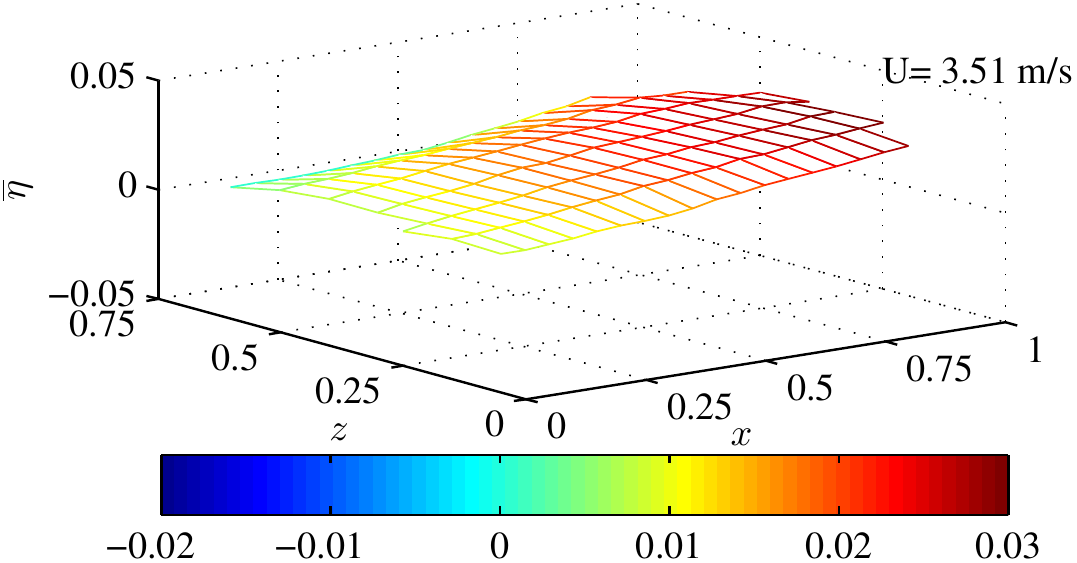}
  \caption{An example of the predominant mean
    component on the surface of a sheet. This particular example
    corresponds to a combined applied tension of T=7.95N acting at
    $\theta$=45\degree to the trailing edge.}
  \label{fig:w0-4004}
\end{figure}

It is worth pointing out that under larger magnitudes of applied
tension at the trailing edge, no noticeable divergence was
observed. Typically, we observe a very small linear increase in the
horizontal displacement of the sheet with a maximum at the trailing
edge of at most 3cm. An example of this in shown in Figure
\ref{fig:w0-4004}. This small mean component is suppressed by higher
fluid velocities and, once $U_c$ is reached, the sheet loses stability
to convected waves before any observable static divergence.

\section{Conclusions}\label{sec:conc}


We have examined the performance of a cantilevered flexible sheet, of
aspect ratio $\order(1)$, mounted horizontally in uniform air flow
with an external in-plane tension applied to the trailing edge. In
particular, an experimental campaign was conducted whereby we
investigated the loss of stability of a flexible sheet to both
convected waves or flutter and static divergence. The unique aspect of
this work was the use of an optical tracking system to measure the
displacement of the sheet surface at select marker locations. We
studied the trajectories of these markers and performed a harmonic
analysis to extract the mode shapes for both divergence and flutter.


Our results have provided a unique insight into the spatial variation
and harmonic composition of an unstable sheet in fluid flow. The
frequency composition of an unstable sheet shows a dominant frequency
component with clearly defined harmonics that suggest the influence of
elastic nonlinearities associated with the sheet are important to
understanding its performance in a post-critical regime. Moreover, we
have shown that a sheet may first lose stability through divergence
before being replaced by a convected wave type instability. During the
energy transfer, we suspect that there is some superposition between
the two instability modes. Under large values of trailing edge
tension, with an equal contribution axially and laterally to the
sheet, no static divergence of the sheet was observed. On the other
hand, a convected wave or flutter type of instability is observed in
this region when stability of the sheet is lost. We have also shown
that, with increasing trailing edge tension, the mode shape amplitudes
of both divergence and flutter are reduced.


Further analysis of these experimental results is required. Of
particular interest is the effect of sheet aspect ratio on the
behaviour of instability modes and also the influence of lateral
tension acting across the sheet. Moreover, we should also like to
examine the frequency shift and phasing of the convected wave
instability across the sheet as the fluid velocity is increased and
the consequences this poses for the fluid dynamic drag. It is the
authors' intention to investigate these aspects further. 

\section*{Acknowledgements}

The authors gratefully acknowledge the financial support of the Norwegian
Research Council under grant number 169417/530 and
Schlumberger. Finally, we thank Rune Toennessen,
Schlumberger, for his comments and technical input over the course of
this work.

\bibliography{/Users/mmthomas/collection/bibliography}
\clearpage

\begin{table*}[t]
  \centering
  \caption{The test matrix employed in the experimental study where $\A$
    denotes the aspect ratio, $A$ the sheet surface area, $m$ the mass
    per unit area, $B$ the flexural rigidity, $\Ren$ the Reynolds
    number, $T$ the combined magnitude of tension applied to the
    trailing edge and $\theta$ represents the 
    angle at which this tension is applied.}
  \begin{center}
  \begin{tabular}{c|c|c|c|c|c|c|c|c}
    sheet &  $\mathscr A$ & A (m$^2$) & $m$ (kgm$^{-2}$)& $B$ ($\times
    10^{-6}$Nm) & $\Ren$ ($\times 10^5$)& $T$ (N) & $\theta = 22.5\degree$
    & $\theta=45\degree$ \\\hline \hline    
    S2 & 1.33 & 0.7500 & 0.1410 & 59.65 &  0.59 - 5.18
    & 2.06 - 7.95 & $\checkmark$ & $\checkmark$ \\ 
    S3 & 1.67 & 0.9375 & 0.1410 & 59.65 &  1.03 - 6.96
    & 2.06 - 7.95 & $\checkmark$ & $\checkmark$\\ 
    S4 & 2.00 & 1.1250 & 0.1410 & 59.65 &  1.01 - 8.17
    & 2.06 - 7.95 & $\checkmark$ & $\checkmark$\\ 
  \end{tabular}
  \end{center}
  \label{tab:parameters}
\end{table*}

\end{document}